\documentstyle{mn}                 %

\def\kms{km\,s$^{-1}$}
\def\mcento{mag\,(100d)$^{-1}$}

\def\Ha{H$\alpha$}

\def\M{M$_{\odot}$}

\input{psfig.tex}

\title[Supernova 1996L]
{Supernova 1996L: evidence of a strong wind episode before the
explosion.
\thanks{Based on observations collected at ESO-La Silla (Chile)}}

\author[Benetti et al.]
{S. Benetti$^{1,2}$, M. Turatto$^{1,3}$, E. Cappellaro$^3$,
I.J. Danziger$^4$, \and P.A.  Mazzali$^4$
\\
$^1$European Southern Observatory, Alonso de Cordova 3107, Vitacura, 
Casilla 19001, Santiago 19, Chile \\
$^2$Telescopio Nazionale Galileo, Apartado de Correos 565, E-38700
Santa Cruz de La Palma, Canary Islands, Spain \\
$^3$Osservatorio Astronomico di Padova, vicolo dell'Osservatorio 5,
I-35122 Padova, Italy \\
$^4$Osservatorio Astronomico di Trieste, via G.B. Tiepolo 11, I-34131 
Trieste, Italy\\
}

\date{Received ................; accepted ................}

\begin{document}

\maketitle

\begin{abstract}
Observations of the type II SN~1996L
reveal the presence of a slowly expanding ($v\sim700$ \kms) shell
at $\sim 10^{16}$ cm from the exploding star.  
Narrow emission features are visible in the early spectra
superposed on the normal SN spectrum. 
Within about two months these features develop narrow symmetric P-Cygni 
profiles.
About 100 days after the explosion the light curve suddenly flattens, 
the spectral lines broaden and the H$\alpha$ flux becomes larger than 
what is expected from a purely radioactive model. 
These events are interpreted as signatures of the onset of the interaction 
between the fast moving ejecta and a slowly moving outer shell of matter 
ejected before the SN explosion.  
At about 300 days the narrow lines disappear and the flux drops 
until the SN fades away, suggesting that the interaction phase is
over and that the shell has been swept away.  
Simple calculations show that the superwind episode started 9 yr before the
SN explosion and lasted 6 yr, with an average $\dot{M}=10^{-3}$ \M/yr.

Even at very late epochs (up to day 335) the typical forbidden lines
of [OI], CaII], [FeII] remain undetected or very weak. 
Spectra after day 270 show relatively strong emission lines of HeI. 
These lines are narrower than other emission lines coming from the SN ejecta, 
but broader than those from the CSM. 
These high excitation lines are probably the result of non-thermal excitation
and ionization caused by the deposition of the $\gamma$-rays emitted in 
the decay of radioactive material mixed in the He layer.

\end{abstract} 
 

\section{Introduction} \label{int}

In a supernova explosion the outer layers of the progenitor star are ejected 
at very large velocities. 
In the case of core-collapse supernovae, which come from massive stars, 
when the ejecta impacts on the circumstellar material (CSM) shed by the 
star during its evolution, a shock front is generated and a fraction of
the kinetic energy of the ejecta is converted into radiation.  
The intensity of the emission depends mainly on the density of the CSM and
on the velocity contrast between the ejecta and the CSM.  
If the density of the CSM is relatively small, the emission from the 
CSM-ejecta interaction becomes visible only after the SN has faded, that is
several years after the explosion. 
If on the other hand the CSM near the SN is very dense, the interaction 
can dominate the SN emission even at early phases.

\begin{figure}
\psfig{figure=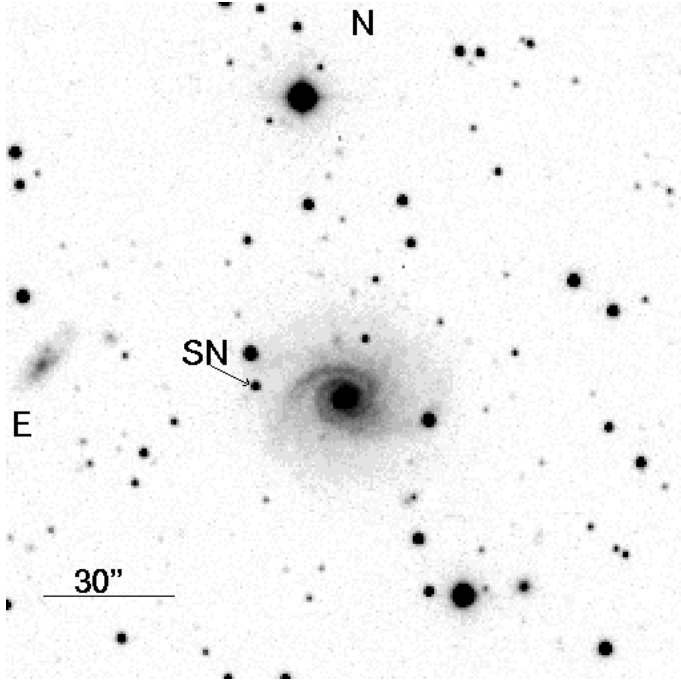,width=9cm,height=9cm}
\caption{SN~1996L in ESO 266-G10. The image is an R frame taken at
ESO3.6m telescope on Mar. 21, 1996. The seeing was 1.1 arcsec}
\label{sn}
\end{figure}

With improved statistics and quality of observations we have now
observed objects fitting in the different scenarios.  
Indeed, in a recent review \cite{macio1} proposed a new classification system 
for type II supernovae (SNII) which is based on the strength of the
CSM-ejecta interaction as determined by an analysis of its signatures.

At one end of the sequence are the classical SNII, where the emission is
determined by thermal balance in the ejecta and the CSM-ejecta
interaction is negligible, at least in the early phases. 
Depending on the shape of their light curve these are divided into plateau 
(II-P) and linear (II-L) SNe. 
The current interpretation is that this subdivision corresponds to a range 
in the mass of the H envelope (from $\sim 10$ \M\/ in IIP to $\sim 1$ \M\/ 
in II-L), which in turn is related to different pre-SN mass loss histories.

At the other extreme are those SNe in which the ejecta interact with
a dense CSM soon after the explosion and the SN light is
by the emission arising from the interaction \cite{terl}.
The best example of this kind of objects is SN~1988Z \cite{macio2}.

In some SNII the evidence of the interaction becomes apparent only when
the SN luminosity fades. 
The fact that, so far, all known objects of this kind are of type IIL is 
consistent with the understanding that SNe IIL experience strong mass loss 
during their evolution.

In an even more rare group of SNe the presence of a dense CSM around the
exploding stars is revealed early-on from the peculiar emission line
profiles, although the ejecta-CSM interaction begins only months later. 
So far, the best representative object of this class (labeled SNIId by Turatto
et al. 1997, where 'd' stands for 'double' profile because of the simultaneous
presence of broad profiles from the ejecta and narrow ones from the CSM) 
was SN~1994aj (Benetti et al. 1997, Paper I).

In this paper we present and discuss the case of SN~1996L, whose photometric 
and spectroscopic properties are very similar to those of SN~1994aj and which 
therefore can be considered a new member of the SNIId class.

\section{Observations} \label{obs}

SN~1996L was discovered by McNaught \shortcite{mc} on Mar. 19.6 in
the outskirts of the galaxy ESO 266-G10 (Fig. \ref{sn}).
The SN was rather faint at discovery ($\sim 18.5$ mag).  
Based on a spectrum obtained two days later \cite{ben1}, the SN
was classified as type II because of the presence of H Balmer lines 
in emission superposed on a blue continuum. 
The Balmer lines had peculiar profiles, showing two emission components, 
one broad and one narrow, both at the rest wavelength of the lines.  
This finding prompted an observational campaign.

\subsection{Photometry} \label{phot}

CCD photometric and spectroscopic observations were obtained on 12
different nights using four different telescopes.  Data reduction
followed the standard procedures described elsewhere (e.g. Paper I).

\begin{figure}
\psfig{figure=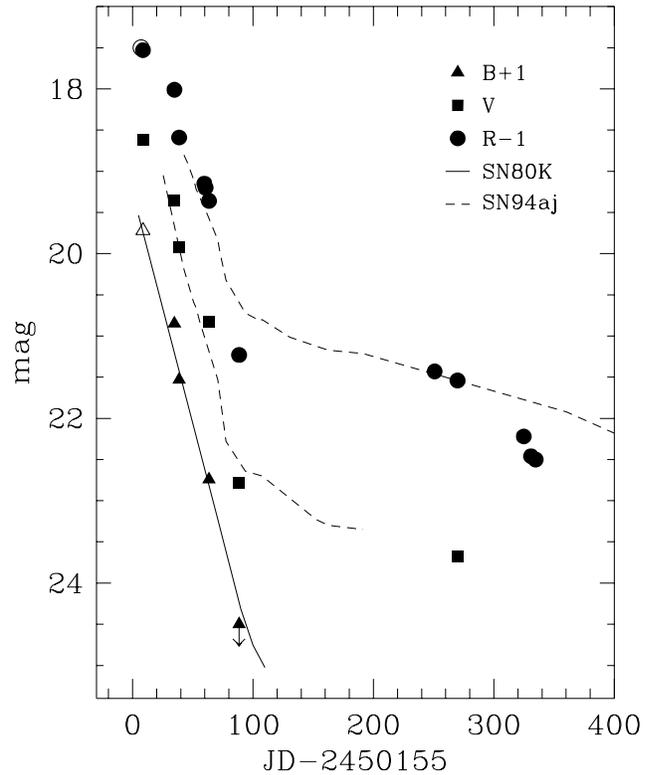,width=9cm}
\caption{B, V and R light curves of SN~1996L. 
The open symbols are the discovery R magnitude (McNaught 1996), and the B 
magnitude derived from our first spectrum. The B and V light curves 
are similar to those of the SNe II 1980K and 1994aj if maximum is assumed 
to have occurred one week before discovery (JD $\sim 2450155\pm 5$d).}
\label{phot_fig}
\end{figure}

The measured magnitudes and the estimated internal
errors are listed in Tab.~\ref{obs_tab}.\\

The B, V and R light curves of SN~1996L are shown in Fig.~\ref{phot_fig}.  
The light curves are clearly of the linear type, although some modulation is 
present during the early decline, as was also the case in other SNe 
(e.g. SN~1994aj).

\begin{table*}
\caption{Photometry of SN~1996L}\label{obs_tab}
\begin{flushleft}
\begin{tabular}{lcccccl}
\hline
    date & J.D.     &  B              &           V     &    R &I&   instr.    \\
         & 2400000+ &                 &                 &       &     &        \\
\hline
 21/3/96 & 50163.7  &  ($18.7\pm0.2$)*  & $18.62\pm 0.02$ & $18.53\pm 0.02$ & &3.6    \\
 16/4/96 & 50189.8  & $19.85\pm 0.04$ & $19.36\pm 0.04$ & $19.01\pm 0.03$ & &3.6 \\
 20/4/96 & 50193.7  & $20.53\pm 0.05$ & $19.92\pm 0.04$ & $19.59\pm 0.04$ & $19.36\pm 0.03$& Dutch  \\
 11/5/96 & 50214.6  &                 &                 & $20.15\pm 0.04$ && NTT  \\
 12/5/96 & 50215.6  &                 &                 & $20.20\pm 0.04$ && NTT   \\
 15/5/96 & 50218.5  & $21.74\pm 0.09$ & $20.83\pm 0.07$ & $20.36\pm 0.06$ &$20.09\pm 0.06$& Dutch \\
  8/6/96 & 50243.5  & $\le23.5$       & $22.78\pm 0.15$ & $22.23\pm 0.10$ && 3.6 \\
18/11/96 & 50405.8  &                 &                 & $22.43\pm0.10$ && Dutch \\
 7/12/96 & 50424.8  &                 & $23.68\pm 0.20$ & $22.54\pm 0.15$ && 3.6 \\
 31/1/97 & 50479.8  &                 &                 & $23.22\pm 0.40$ && Dutch \\
  6/2/97 & 50485.8  &                 &                 & $23.46\pm  0.20$& & 3.6 \\
9-10/2/97& 50490.0  &                 &                 & $23.50\pm0.40$ && 3.6 \\
\hline
\end{tabular}

$*$ derived from spectrophotometry

3.6 = ESO 3.6m telescope + EFOSC1

NTT = ESO NTT + EMMI

Dutch = Dutch 0.90m + CCD Camera
\end{flushleft}
\end{table*}

As a measure of the overall luminosity decline we use the parameter
${\beta}_{100}$ \cite{pat1}, obtaining the values
${\beta}^B_{100}\ge~6.0$, ${\beta}^V_{100}=5.2$ \mcento.  These values
are similar to those of the SNe II-L 1980K and 1994aj, whose light
curves are also shown in Fig.\ref{phot_fig}.
Since the light curves of SN~1996L appear to be very similar to those of the 
other two SNe, although they may not be identical, in the following we will 
adopt as reference epoch $JD \sim 2450155$ (Mar 12, 1996), which gives the 
best fit of the light curves of SN~1996L with those of SN 1980K and 1994aj. 
This choice leads to an estimate of the time of maximum, which should have 
occurred about a week before discovery, with an uncertainty of $\pm 5$ days.  
We also estimate that the SN reached a maximum magnitude
$B_{max}\sim V_{max} \sim 18.5\pm0.2$.
 
Unfortunately observations were interrupted because of the seasonal gap, but 
apparently the luminosity decline slowed down significantly between 90 and 270 
days (${\gamma}_{\rm R}=0.15$ \mcento).  
After day 270, the R light curve decreased rapidly again, with
${\gamma}_{\rm R}=1.4$ \mcento.

SN~1996L suffers moderate galactic reddening, E(B-V)$=0.07$ \cite{bh}
and there is no evidences of additional extinction in the parent galaxy.

The heliocentric velocity of the parent galaxy is $9900\pm 100 km~s^{-1}$ 
as measured from the HII emission lines visible in the long slit spectra. 
This corresponds to $\mu=35.60-5logH_0/75$.

Correcting for the extinction, we obtain $M^0_B \sim -17.4-5logH_0/75$. 
This is very close to the absolute magnitude at maximum of SN~1994aj 
($M^0_B = -17.8$, Paper I), and lies between the average values 
for `regular' and `bright' SNe II-L ($< ~M^0_B~ > = -16.8 \pm 0.5$ and 
$< ~M^0_B~ > = -18.9 \pm 0.6$, respectively, for $H_0=75$, Patat et al. 1993).

The main data of SN~1996L and those of its parent galaxy are
summarized in Tab. \ref{data}.

\subsection{Spectroscopy} \label{spec}

The journal of the spectroscopic observations is given in Tab.\ref{spec_tab}.  
The table lists for each spectrum the date (col.1), the phase (col.2), the 
equipment used (col.3), the exposure time (col.4), the wavelength range (col.5),
and the resolution, which is the measured FWHM of the night-sky lines (col.6). 
In order to improve the S/N ratio, in some cases exposures taken on 
different nights have been averaged. 
In those cases we list the cumulative exposure time.

\begin{table}
\caption{Main data of SN~1996L}\label{data}
\begin{tabular}{ll}
\hline
Parent galaxy 		& ESO 266--G10 (PGC 36065)	\\
Galaxy type		&	Sa:	$^\dag$		\\
RA (2000)  		&  $11^h 38^m 38^s.80$        		\\
Dec (2000) 		& $-43\degr$25'11''.6  		\\
Recession velocity	[\kms] & $9900\pm 100$ 	\\
Distance modulus (H$_0=75$)	& 35.60			\\
Galactic A$_B$		&	0.30$^\ddag$ 		\\
Offset from nucleus 	& 	29".5E  ~~  2".2N       \\				
			&				\\
Date of maximum		& JD $\sim 2450155 \pm5$ (Mar 12, 1996) \\
magnitude at max	& $B_{max}\simeq V_{max}\simeq18.5\pm0.2$ \\
decline rate [\mcento] & $\beta^B_{100}\ge 6.0$,  $\beta^V_{100}=5.2$ \\

\hline
\end{tabular}

{\dag} NED

{\ddag} Burnstein \& Heiles ApJS 54, 33 (1984)
\end{table}

\begin{table}
\caption{Spectroscopic observations of SN~1996L} \label{spec_tab}
\begin{tabular}{lrcrrr}
\hline
\hline
     Date     & phase$^*$ & inst.$^{**}$  &  exp  &   range   
	 & res.\\
              & (days)&       & (min) &   (\AA)   &    (\AA)   \\
\hline
     21/3/96  &  +9   &  3.6  &  20   & 3750-6950 &     15     \\
     16/4/96  &  +34  &  3.6  &  30   & 3750-6950 &     15     \\
     25/4/96  &  +44  &  1.5  & 180   & 3060-10470&     13     \\
  11-12/5/96  &  +61  &  NTT  & 120   & 3640-8940 &     13     \\
     7/12/96  & +270  &  3.6  &  20   & 6000-9800 &     15     \\
  5-9-10/2/97 & +335  &  3.6  & 240   & 3750-9800 &     15     \\
\hline
\end{tabular}

* - relative to the estimated epoch of maximum, JD=2450155

** - See note to Table~1 for coding. 1.5 = ESO 1.5m + B\&C
\end{table}

The flux calibration of the spectra was checked against the photometry and
in case of disagreement the spectra were adjusted.

\begin{figure*}
\psfig{figure=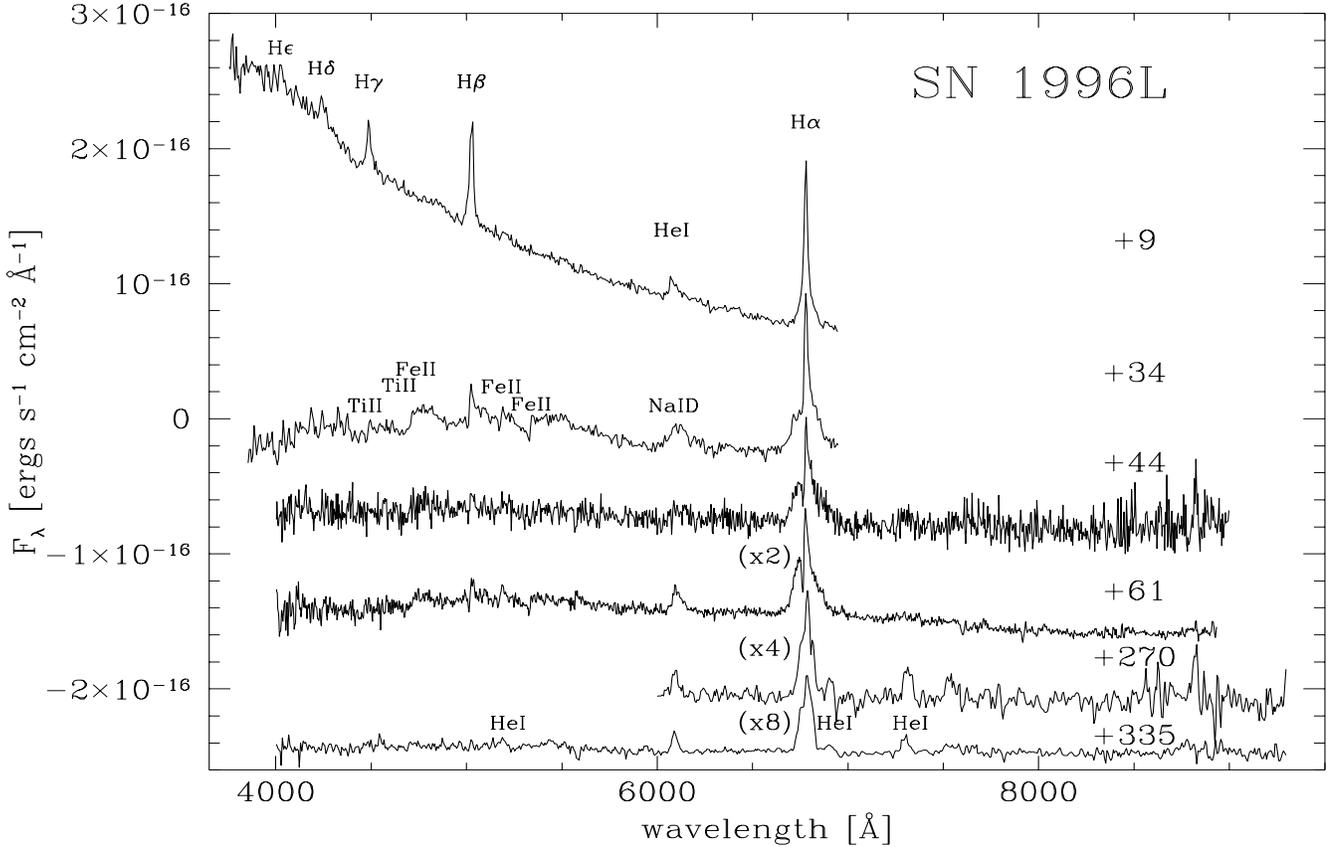,width=18cm,height=12cm,angle=270}
\caption{Spectral evolution of SN~1996L. Wavelength is in the
observer rest frame. The ordinate refers to the first spectrum (+9d),
the other spectra are shifted downwards by $0.7\times 10^{-16}$,
$1.0\times 10^{-16}$, $1.7\times 10^{-16}$, $2.1\times 10^{-16}$ and
$2.5\times 10^{-16}$ respectively. For clarity the last three spectra
have been multiplied by a factor of 2, 4 and 8 respectively.}
\label{spec_fig}
\end{figure*}

Figure \ref{spec_fig} illustrates the spectroscopic evolution of
SN~1996L from phase +9d to +335d.

The distinctive features of SNII, the HI Balmer lines, dominate the
spectrum of SN~1996L at phase +9d. At this epoch, the spectrum shows a
very blue continuum (Tbb $\sim 16000$K with the adopted reddening).
The Balmer lines have two components, a relatively broad emission and a 
superposed narrow component.  As in SN~1994aj, 
the Balmer lines never show any evidence of broad P-Cygni absorption.  
Generally this is attributed to the fact that net H$\alpha$ emission caused 
by the collisional excitation of H \cite{branch} obliterates the P-Cygni 
absorption originating from the redistribution of the photospheric continuum.
This effect is usually strongest in H$\alpha$, because of the lower energy 
required compared, for instance, to H$\beta$, and because of the larger 
optical depth of this line.  
In the case of SN~1996L, however, H$\beta$, H$\gamma$, and H$\delta$ 
do not show broad P-Cygni absorptions either.  
As for other SNII, we believe that the narrow component arises from the CSM
excited by the hard UV and X--ray flash emitted at shock break-out.

\begin{table*}
\caption{Spectral lines characteristics of SN~1996L} \label{lines}
\begin{flushleft}
\begin{tabular}{cccccccccccccc}
\hline
\hline
 phase &                     &        & H$\alpha$ &      & & &H$\beta$  &      && &H$\gamma$ &  & HeI$-$NaID \\
\cline{3-5}\cline{7-9}\cline{11-13}
 (days)&                     & Broad  & (Em)$_n$  & (Ab)$_n$&& Broad &(Em)$_n$  & (Ab)$_n$& & Broad& (Em)$_n$&(Ab)$_n$ &   broad \\
\hline
       & $\lambda_c$ (\AA)   & 6779.5 & 6783      &      & &  5024  
       &
5028   &      &&  4497: &  4487 &      &  6076:  \\
  +9   & FWHM (\AA)          &  80    &  19       &      & &   44
  &  18
&      &&   50:  &  16     &      &  48:    \\
       & $\phi$$^*$&  26    &  18       &      & &   12    &
10       &      &&    6:  &   4     &      &   6:    \\
\hline
       & $\lambda_0$ (\AA)   & 6789.5 & 6780      &      & &  5069
       & 5025
& 5004:&&        &  4495   & 4473 &  6110    \\
  +34  & FWHM (\AA)          & 133    &  18       &      & &   80
  &
&      &&        &         &      &   95     \\
       & $\phi$ $^*$&  56    &  15       &      & &   8.5   &
4.8      &  0.6 &&        &   1.8   & 0.7  &   18     \\
\hline
       & $\lambda_0$ (\AA)   & 6786   & 6781      & 6765 & &
       & 5031
& 5007 &&        &         &      &         \\
  +44  & FWHM (\AA)          & 150    &           &      & &
  &
&      &&        &         &      &         \\
       & $\phi$$^*$& 55.5   & 6.5       & 2.4  & &         &
3.4:     & 0.6: &&        &         &      &         \\
\hline
       & $\lambda_0$ (\AA)   & 6777   & 6779      & 6764 & &  5049
       &
5031     & 5013 &&        &         &      & 6095    \\
  +61  & FWHM (\AA)          & 135    &   15        & 13     & &
  55    &
&      &&        &         &      &  54     \\
       & $\phi$  $^*$&  32    & 2.8       & 1.6  & &   5.0   &
1.0     &  0.6 &&        &         &      &   5     \\
\hline
       & $\lambda_0$ (\AA)   & 6781   & 6787      &      & &
       &
&      &&        &         &      & 6091    \\
  +270 & FWHM (\AA)          &  83    &  16       &      & &
  &
&      &&        &         &      &  32     \\
       & $\phi$$^*$ & 12    & 1.5       &      & &         &
&      &&        &         &      &  2.0    \\
\hline
       & $\lambda_0$ (\AA)   & 6782   & 6787      &      & &
       &
5026:    &      &&        &         &      & 6091    \\
  +335 & FWHM (\AA)          &  83    &  21       &      & &
  &   17
&      &&        &         &      &  35     \\
       & $\phi$$^*$&  4.0   &  0.6      &      & &         & $\le
0.2$ &      &&        &         &      &  0.7    \\
\hline

\end{tabular}

$*$ Flux is in units of erg\,s$^{-1}$\,cm$^{-2} \times 10^{-16}$
\end{flushleft}
\end{table*}

As time progresses the spectrum of the SN becomes rapidly cooler, and the 
narrow H$\alpha$ emission develops an absorption component, until two months
after maximum it displays a symmetric P-Cygni profile (cf. Sect.~3). 

\begin{figure}
\psfig{figure=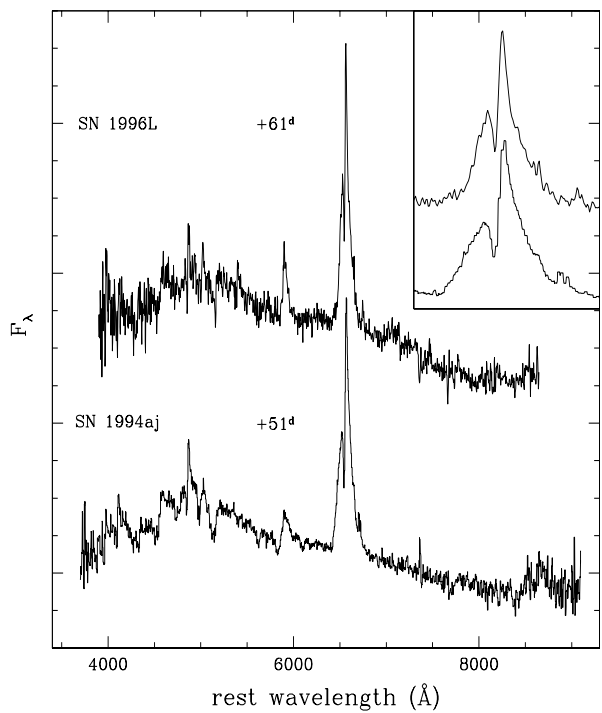,width=9cm,height=10cm}
\caption{Comparison of the photospheric spectrum of SN~1996L with that
of SN~1994aj (Paper I). Wavelength is in the galaxy rest frame
and the spectra have been corrected for galactic extinction. 
The panel at the top-right shows an enlargement of the \Ha~profiles.}
\label{fig_phot}
\end{figure}

In Fig. \ref{fig_phot} we compare the spectrum of SN~1996L at a phase
+61d with that of SN~1994aj at a phase +51d. 
The spectra are very similar to one another, suggesting that the presupernova 
histories of the two objects were analogous.  
The greatest difference is the lack of the broad NaID 
absorption in the spectrum of SN~1996L.

The FeII and TiII lines show the usual P-Cygni profiles, from which an 
expansion velocity of about 1700 \kms is derived. 
This is smaller than the velocity derived for SN~1994aj (about 2500 \kms).

About one year after maximum H$\alpha$ was still the dominant feature
in the spectrum of SN~1996L. As in SN~1994aj, forbidden lines are very
weak (CaII] 7291-7323 \AA), or absent.  However, unlike SN~1994aj, the
spectra of SN~1996L at phases 270 and 335 days show emission lines of
HeI 5015, 5876 (blended with NaID), 6678, and 7065 \AA.  The expansion
velocity deduced from the FWHM of the HeI lines is about 1700 \kms.
The 7065 \AA\ line is remarkable for its strength when compared to
other HeI lines such as 5876 which, according to recombination theory,
should be stronger.

\section{H$\alpha$ profile and flux evolution} \label{halpha}

\begin{figure}
\psfig{figure=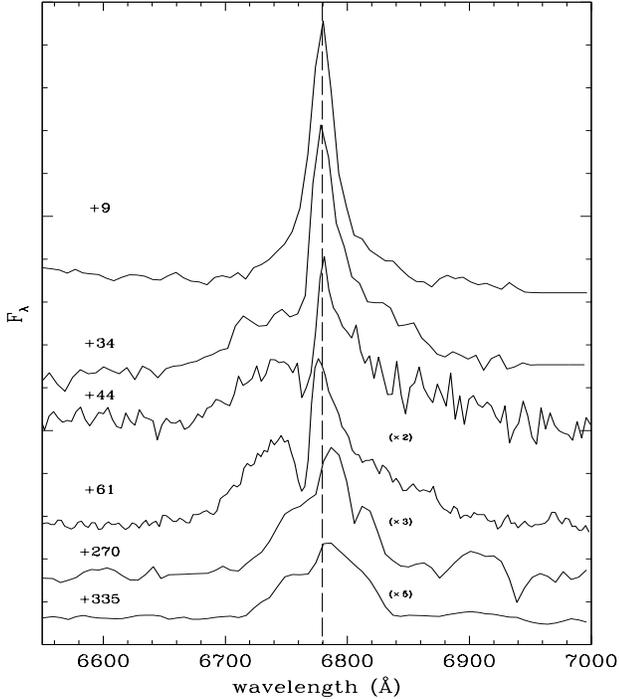,width=9cm,height=10cm}
\caption{Evolution of H$\alpha$ profile in SN~1996L. 
The long dashed line marks the galaxy rest frame position of the transition. 
In order to enhance the contrast the last three spectra have been multiplied 
by the factors shown in the figure.}\label{Ha_profile}
\end{figure}

It is interesting to study the evolution of the H$\alpha$ line.
Figure~\ref{Ha_profile} shows an enlargement of the spectra of
SN~1996L and Table~\ref{lines} reports the main line parameters
derived from multiple Gaussian fitting (using the ALICE package in MIDAS).

The H$\alpha$ profile at 9d consists of a broad, symmetric emission
component ($FWHM \sim~3500$ \kms) and a narrow emission superposed to it 
and centred at the same wavelength which is barely resolved 
($FWHM \sim~500$ \kms, after correction for instrumental resolution).  
The profiles of H$\beta$ and H$\gamma$ are similar to that of H$\alpha$.

The H$\alpha$ profile undergoes major modifications with time. 
The broad component becomes progressively broader, until at phase 61d it
reaches a $FWHM \sim~6000$ \kms. Because of the high S/N ratio of this
spectrum, a small asymmetry in the broad component is revealed.  
The red-wing zero intensity velocity of $\sim~6900$ \kms\ is in fact
somewhat larger than the corresponding blue-wing velocity, $\sim~5500$ \kms.  
Such an asymmetry, which was also observed in SN~1994aj at a similar phase, 
could be explained if the scattering optical depth of H$\alpha$ photons is 
significant. 
Electron scattering is most probably responsible for this large optical depth. 
The expansion velocity on SN~1996L is smaller than in SN~1994aj and
the H$\alpha$ asymmetry is less pronounced.  
Also, the narrow emission becomes more and more asymmetric with time, 
and between phases 44d and 61d it turns into a P-Cygni profile 
(the intensity of the absorption being similar to that of the emission).  
The blue wing of the absorption indicates a maximum velocity of 
$\sim 1600$ \kms, whereas the minimum of the trough corresponds to 
$\sim 700$ \kms, which is comparable to the velocity of the CSM emission 
lines at early phases.  
The profile of the narrow component is very similar to that of SN~1994aj 
(see Fig. \ref{fig_phot}), but slightly narrower: the velocities in SN~1994aj 
were 2000 \kms for the blue wing and 900 \kms for the minimum of the 
absorption (Paper I).

It is noteworthy that a narrow P-Cygni feature appears also in H$\beta$. 
Although the S/N is not good enough to analyze the profile of H$\beta$ in 
detail, the expansion velocity is consistent with that of H$\alpha$.

As for SN~1994aj, we believe that the narrow component originates in the CSM,  
which is excited by the hard radiation emitted at shock break-out.  
The CSM is probably the result of a strong wind episode taking
place in the SN progenitor some time before the explosion.

When we recovered the SN, at phase 270d, the narrow feature had lost its 
P-Cygni profile and consisted of an unresolved emission line of much lower 
intensity. The FWHM of the broad emission had decreased to a velocity of 
$\sim~3700$ \kms. By this time we 
suggest that the bulk of the interaction between the SN ejecta and the CSM, 
which started about 100 days after maximum (see Sec. \ref{phot}), is over.
This is also suggested by the R and H$\alpha$ light curves (see below). 
At these phases the H$\alpha$ profile is quite different from that of 
SN~1994aj at similar phases. 
In SN~1994aj the broad component had a flat-top profile, which is the 
signature of a shell, interspersed with narrow absorptions (see Paper I).

\begin{figure}
\psfig{figure=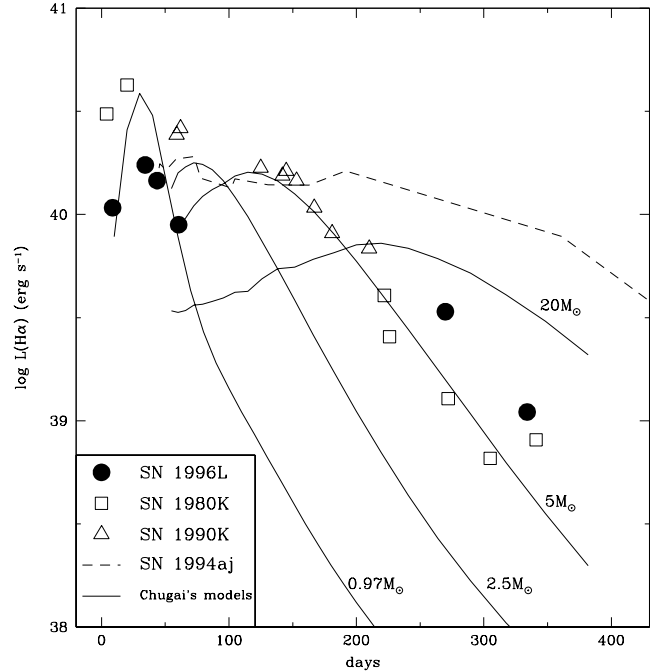,width=10cm,height=10cm} 
\caption{Evolution of the \Ha~ luminosity of SN~1996L compared with
the same data for SNe 1980K (Uomoto \& Kirshner, 1986), 1990K
(Cappellaro et al. 1995b), and SN~1994aj (Paper I). 
Also plotted are Chugai's models (Patat et al. 1995; Chugai, 1998, private
communication).}
\label{Halpha-flux}
\end{figure}

In Fig.~\ref{Halpha-flux} we show the absolute \Ha\ light curve of SN~1996L 
(see also Tab. \ref{lines}) and compare it with those of the type II-L 
SNe~1994aj, 1990K and 1980K. The H$\alpha$ flux in SN~1996L reached a maximum
around phase +34 days, during the line broadening phase, when the narrow
component consisted only of pure emission.  The H$\alpha$ luminosity
of SN~1996L is similar to that of SN~1994aj up to about 50 days after
maximum, but at later phases SN~1996L declines much more rapidly.

\section{Discussion}

In the previous sections we have seen that SN~1996L shares many of the 
properties of another SNII with a double P-Cygni profile, SN~1994aj, which 
was discussed in Paper I. 
Since SN~1996L was discovered soon after maximum it was possible to follow 
the evolution of the strong narrow \Ha\ emission into a P-Cygni shaped
line and then its progressive disappearance. 
We believe that the narrow Balmer emission lines are produced by the 
recombination of the CSM ionized by the UV and X-ray flash emitted at shock 
break-out.  
This is the same mechanism as for the narrow lines observed in SN~1993J in 
the first few days after the explosion \cite{ben93j}. The emission
intensity in SN~1996L is two orders of magnitude larger but coronal lines of
[FeX] and [FeXIV] were visible in SN~1993J while they are not in SN~1996L.  
When the recombination is sufficient for the Sobolev optical depth of \Ha\ 
in the slow-moving CSM shell to be significant (typically values larger 
than one are required in at least some velocity shells), the shell can scatter 
the continuum radiation and the emission line takes on a P-Cygni profile. 
In SN~1996L this happened about 50 days after maximum. 
It should be noted that for significant recombination to occur on this time 
scale the density of the CSM must be relatively high, 
$n_{\rm e} > 10^5 cm^{-3}$.
 
The P-Cygni profile is narrow because the CSM has a low velocity. 
The velocity of the most highly ionized gas as measured in the early 
emission lines ($FWHM~\sim~500$ \kms), is similar to that of the minimum 
of the P-Cygni profile ($\sim 700$ \kms), suggesting that both features 
originate in the same location.

The early light curve of SN~1996L shows the fast decline typical of SNe II-L, 
which is indicative of a small envelope mass at the time of explosion.  
This is thought to be the result of large mass loss during the evolution of 
the progenitor which is also the origin of the dense CSM.

About three months after maximum, when other SNII enter the so-called 
radioactive tail with a decline rate of about $\gamma \simeq1$ \mcento\ 
\cite{macio}, the decline of the light curve of SN~1996L suddenly slows 
down, to a rate of ($\gamma_R=0.15$ \mcento). 
Since this is much slower than the $^{56}$Co decay rate, an additional 
energy input is clearly required to power the light curve.  
In analogy with the case of SN~1994aj (Paper I), we propose that the extra 
energy comes from the interaction of the SN ejecta with the dense CSM.  
At around 300 days the decline rate increases again (${\gamma}_{\rm R}=1.4$ 
\mcento), suggesting that at this epoch the ejecta has swept away most of the 
circumstellar shell and that the bulk of the interaction is over.
After this sharp decline marking the end of the interaction phase, the light 
curve should finally settle on the radioactive tail, but because of the 
distance of the SN this last transition was not observable.

Given the maximum expansion velocity of the ejecta (6900 \kms), and 
assuming that the interaction began on day 100, an inner radius of 
$r_{\rm i} \sim 6\times 10^{15}$ cm is derived for the pre-SN wind. 
If the wind velocity was 700 \kms, as derived from the width of the narrow 
lines, this means that the strong wind episode terminated about 3 years 
before the explosion. 
If the bulk of the interaction ended at a phase of about 300 days, the outer
edge of the wind has a radius $r_{\rm e} \sim 2 \times 10^{16}$ cm.
This implies that the strong wind episode lasted about 6 years.
If we assume that the wind was steady, with constant mass loss rate $\dot{M}$ 
and terminal outflow velocity $V_\infty$, and thus produced an $r^{-2}$ 
density profile, we can obtain a lower limit for the mass in the CS shell
(see Paper I):

$$ M \ga 4\pi r_{\rm e}^3 m_{\rm H} n_{\rm e}(r_{\rm e})
 (1-\frac{r_{\rm i}}{r_{\rm e}}) \sim 0.01 M_{\odot}$$

This is an order of magnitude smaller then the estimated mass of the CS shell
in SN~1994aj (Paper I).

If we assume that the mass loss was constant for a period of about 6 years, we
obtain a mass loss rate of $\sim 1.5 \times 10^{-3}$ M$_{\odot}$/yr for
SN~1996L. 

We can also obtain a rough estimate of the wind density and the mass 
loss rate from the observed H$\alpha$ luminosity during the interaction phase 
using the expression

$$ L(H\alpha) = \frac{1}{2} \psi \frac{\dot{M}}{u_{\rm w}} v^3_{\rm sn}$$

\noindent
where $\psi$ is the efficiency with which the kinetic energy is converted into
H$\alpha$ radiation, $\dot{M}$ is the mass loss rate in M$_{\odot}$/yr, 
$u_{\rm w}$ the wind velocity and $v_{\rm sn}$ the velocity at the CSM-wind 
contact discontinuity. 
The value of $\psi$ is not well known, but it should be of the order of
$0.1$ \cite{chu1}.  
If we adopt for SN~1996L a mean value of $v_{\rm sn} \simeq4800$ \kms for 
$v_{\rm sn}$, then from L(\Ha)=4.47$\times 10^{39}$ erg s$^{-1}$
we obtain 
$\dot{M}/u_{\rm w} \sim 8 \times 10^{14}$ g/cm, which is half the value 
obtained in Paper I for SN~1994aj.  
Since the velocity of the CS shell of SN~1996L is about 700 \kms, then 
the mass loss rate should be $\dot{M} \sim 9 \times 10^{-4}$ M$_{\odot}$/yr.
This is roughly consistent with the value derived above.

Fig.~\ref{Halpha-flux} shows that after day 30 the \Ha\ luminosity 
of SN~1996L is significantly smaller than that of SN~1994aj.  
After day 270, the \Ha\ luminosity decline of SN~1996L is comparable to the 
\Ha\ light curves computed by Chugai (Patat et al. 1995; Chugai, 1998, private
communication) assuming that \Ha\ at late times is powered only by the
radioactive decay of Co to Fe.  A comparison of the models with the
observations in the first 2-3 months, before the beginning of the
interaction phase, suggests that the ejecta mass is $<2.5$
M$_{\odot}$, and most likely $\simeq 1$ M$_{\odot}$.  The last two
\Ha\ points appear to follow a model for an ejecta mass in excess of
5 M$_{\odot}$, but they are probably still affected by the
interaction.  Observations at later epochs, which are beyond the
limits of 4-m class telescopes, would be necessary to establish
conclusively the mass of the ejecta.

Another interesting difference between SN~1996L and SN~1994aj is that
strong He lines are present in the spectra of SN~1996L at late phases. 
In our last spectrum the ratio of the 5876 and the 7065\AA\ line is $<1.2$. 
This is only an upper limit because the 5876\AA\ line could be
blended with the NaID lines. 
Since in LTE conditions the ratio should be 5, and the temperature at such 
epochs is too low for He to be photoionized, non--thermal excitation could be
responsible for the very non-LTE conditions which must be invoked to explain 
both the presence of the HeI lines and their unusual ratios. 

We remarked above that at about 300 days the ejecta-CSM interaction is
essentially over.  
Therefore, the most probable cause of these non-LTE effects could be the 
mixing in the He mantle of radioactive material synthesized in the explosion, 
mainly $^{56}$Ni which decays into $^{56}$Co. 
$^{56}$Ni dredge-up has been already seen in other SNII, 
e.g. SN~1987A \cite{graham}, and SN~1995V \cite{fassia}. 
Since the envelope mass of SN~1996L is probably very small, at 300 days 
the envelope should be sufficiently thin that inner regions are exposed. 
Moreover, the ratio of the strengths of the broad H$\alpha$ component and 
of HeI 6678 \AA\ is sensitive mostly to the He abundance. 
At phase +335d this ratio is about 10, which is much smaller than the
normal ratio of about 100 which occurs for a normal He abundance. 
This finding reinforces the hypothesis that by this time we are seeing deep 
into the SN ejecta, near the He-rich envelope.  
This hypothesis is supported by the fact that the He lines are broader 
(1700 \kms) than the CS shell lines (FWHM$\sim700$ \kms), but narrower than 
those from the shocked ejecta (FWHM$\sim 3700$ \kms).  
Therefore the He lines could arise from the inner, slower ejecta.

\section{Conclusions}

The observations presented in this paper demonstrate that SN~1996L belongs 
to the newly defined class of type IId SNe, whose prototype is SN~1994aj. 
SNIId are characterized by showing signatures of an interaction between the 
SN ejecta and CSM ejected by the progenitor star. 
The interaction starts a few months after the explosion.

Another common property of SNIId is that, in spite of the considerable CSM 
mass detected around these objects and the implied large mass loss prior 
to explosion, the [OI] 6300-6364 \AA~ lines are not detected. 
This strongly suggests that the progenitors of these supernovae had main 
sequence masses smaller than $\sim 8$M$_{\odot}$ \cite{cd}.

Interacting SNII show a wide range of properties, as summarized in a recent 
paper by Chugai \shortcite{chu2}. 
The two main parameters defining the observed phenomena are the pre-SN mass 
loss rate and the wind characteristics (uniform or clumpy). 
Interacting SNe include SNe IIn, SN~1979C and other SNe II-L recovered long
after the explosion (such as SN~1980K). 
Along this sequence the explosions appear to have occurred in a increasingly 
less dense CSM, suggesting a trend towards smaller mass loss rates by the 
progenitor stars \cite{chu2}.\\ 
In particular, Type IIn supernovae (called IIdw in Chugai's paper) are 
thought to occur inside a dense wind ($\dot{M}>10^{-4} u_{10}$ M$_{\odot}$/yr, 
where $u_{10}$ is the wind velocity in units of 10 \kms). 
The wind can be either clumpy (eg. SN~1988Z) or uniform (eg. SN~1987F). 
Interaction masks the thermal emission from the ejecta.  
Less extreme cases are those of SN~1979C, whose pre-SN wind was uniformly 
distributed and had a mass loss rate of 
$\dot{M}\sim 1.2\times 10^{-4} u_{10}$ M$_{\odot}$/yr, 
or SN~1980K where the wind has only a moderate density, 
$10^{-5} u_{10} < \dot{M} < 10^{-4} u_{10}$ M$_{\odot}$/yr.

Additional parameters seem to be required to describe the wide range
of properties displayed by the class of SN~IId, which are the epoch
and the duration of the last strong wind episode, relative to the time
of explosion.

If the pre-SN wind lasted for many years and continued almost until
explosion, the interaction starts soon after the SN event and
outshines the SN itself, giving rise to a SNIIn.  If on the other hand
the pre-SN wind episode ends a few years before the explosion, then
the supernova is of type IId and has the properties described in this
paper for SN~1996L.  Finally, if the pre-SN wind ended more than about
a hundred years before the explosion, the material in the wind shell
is so diluted by the time the SN ejecta can reach it that the
observable signatures of the interaction are very weak and may not be
detected as is the case for some SNII-L and for SNII-P.

\bigskip

\noindent
{\bf ACKNOWLEDGMENTS} We are grateful N.N. Chugai for communicating his 
\Ha light curve results prior to publication.  
SB would also like to thank R. Terlevich for stimulating discussions.


\noindent

\begin{thebibliography}{cosaservira}     
\bibitem[\protect\citename{Benetti et al. }1994]{ben93j}
        Benetti, S., Patat, F., Turatto, M., Contarini, G., Gratton,
        R., Cappellaro, E., 1994, A\&A, 285, L13
\bibitem[\protect\citename{Benetti et al. }1996]{ben1}
        Benetti, S., Barthel, P., de Vries, W., 1996, IAUC 6346
\bibitem[\protect\citename{Benetti et al. }1997]{p1}
        Benetti, S., Cappellaro, E., Danziger, I.J.,
        Turatto, M., Patat, F. Della Valle, M., 1997,
        MNRAS., in press (Paper I)
\bibitem[\protect\citename{Branch et al. }1981]{branch}
        Branch, D., Falk, S.W., McCall, M.L., Rybski, P., Uomoto,
        A.K.,
        Wills, B.J., 1981, ApJ 244, 780
\bibitem[\protect\citename{Burstein \& Heiles }1984]{bh}
        Burstein, D., Heiles, C., 1984, ApJS 54, 33
\bibitem[\protect\citename{Cappellaro et al. }1995a]{cap95}
        Cappellaro, E., Danziger, I.J., Turatto, M., 1995a, MNRAS 277,
        106
\bibitem[\protect\citename{Cappellaro et al. }1995b]{capp}
        Cappellaro, E., Danziger, I.J., Della Valle, M., Gouiffes, C.,
        Turatto, M., 1995b, A\&A 293, 723
\bibitem[\protect\citename{Chugai }1990]{chu1}
        Chugai, N.N., 1990, SA 16, L457 AJ 111, 1286
\bibitem[\protect\citename{Chugai }1997]{chu2}
        Chugai, N.N., 1997, Astronomy Reports, 41, 672
\bibitem[\protect\citename{Chugai \& Danziger }1994]{cd}
        Chugai, N.N., Danziger I.J., 1994, MNRAS 268, 173
\bibitem[\protect\citename{Fassia et al. }1998]{fassia}
        Fassia, A., Meikle, W.P.S., Geballe, T.R., Walton, N.A., Pollacco,
        D.L., Rutten, R.G.M., Tinney, C., 1998, MNRAS 299, 150
\bibitem[\protect\citename{Graham }1988]{graham}
        Graham, J.R., ApJ 335, L53
\bibitem[\protect\citename{McNaught }1996]{mc}
        McNaught, R.H., 1996, IAUC 6346
\bibitem[\protect\citename{Patat et al. }1993]{pat1}
        Patat, F., Barbon, R., Cappellaro, R., Turatto, M., 1993,
        A\&A 282, 731
\bibitem[\protect\citename{Patat et al. }1995]{pat}
        Patat, F., Chugai, N., Mazzali, P.A., 1995, 
        A\&A 299, 715
\bibitem[\protect\citename{Terlevich }1994]{terl}
        Terlevich, R.J., 1994, {\it Circumstellar Media in the Late
        Stages of Stellar Evolution} eds. R.E.S. Clegg, I.R. Stevens,
        W.P.S. Meikle, Cambridge Univ. Press, Cambridge, p. 153
\bibitem[\protect\citename{Turatto et al. }1990]{macio}
        Turatto, M., Cappellaro, E., Barbon, R., Della Valle, M.,
        Rosino,L., 1990, AJ 100, 771
\bibitem[\protect\citename{Turatto et al. }1993]{macio2}
        Turatto, M., Cappellaro, E., Danziger, J., Benetti, S.,
        Gouiffes, C., Della Valle, M., 1993, MNRAS 262, 128
\bibitem[\protect\citename{Turatto et al. }1997]{macio1}
        Turatto, M., Benetti, S., Cappellaro, E., Danziger, I.J.,
        Mazzali, P.A., 1997, in {\it SN 1987A: Ten years After}, Fifth
        CTIO/ESO/LCO Workshop, eds. Phillip, M.M. and Suntzeff, N.B.,
        in press
\bibitem[\protect\citename{Uomoto \& Kirshner }1986]{uk}
        Uomoto, A., Kirshner, R.P., 1986, ApJ 308, 685
\end{thebibliography}
\end{document}